\documentstyle[aps,12pt]{revtex}

\begin{document}

\title{Exact solutions to the Dirac equation\\
for a Coulomb potential in $D+1$ dimensions}

\author{Xiao-Yan Gu \thanks{Electronic address:
guxy@mail.ihep.ac.cn} and Zhong-Qi Ma \thanks{Electronic address:
mazq@sun.ihep.ac.cn} }

\address{Institute of High Energy Physics, Beijing 100039,
The People's Republic of China}

\author{Shi-Hai Dong \thanks{E-mail address: dongsh@nuclecu.unam.mx}}

\address{Instituto de Ciencias Nucleares, UNAM,
Apartado Postal 70-543\\
Circuito Exterior, C. U. 04510 Mexico, D. F. ,  Mexico}


\maketitle

\vspace{5mm}

\begin{abstract}

The Dirac equation is generalized to $D+1$ space-time. The
conserved angular momentum operators and their quantum numbers are
discussed. The eigenfunctions of the total angular momenta are
calculated for both odd $D$ and even $D$ cases. The radial
equations for a spherically symmetric system are derived. The
exact solutions for the system with a Coulomb potential are
obtained analytically. The energy levels and the corresponding
fine structure are also presented.

\vspace{2mm}

\noindent


\end{abstract}

\section{Introduction}

The exact solutions of the Schr\"{o}dinger equation in the real
three-dimensional space for a hydrogen atom and for a harmonic
oscillator were important technical achievements in quantum
mechanics \cite{sch,dir}. During the past half century, the
mathematical tools for the orbital angular momentum operators and
their eigenfunctions in an arbitrary $D$-dimensional space have
been presented \cite{erd,lou1,lou2,lou3,cha,ban}. The
nonrelativistic $D$-dimensional Coulombic and the harmonic
oscillator problems have been studied in some detail by many
authors \cite{wod,ben1,ben2,rom,alj1,lin,hos}. The solutions of
the Dirac equation, however, have been studied in the usual three-
\cite{kol,hyl,mar,auv,wal}, two- \cite{don} and one-dimensional
\cite{lap} space. Motivated by the recent interest of
higher-dimensional field theory , we generalize the Dirac equation
to $D+1$ space-time. The conserved total angular momentum
operators and their quantum numbers are discussed. The
eigenfunctions of the total angular momenta are calculated for
both odd $D$ and even $D$ cases. From the viewpoint of
mathematics, this problem is a typical application of group theory
to physics. In terms of the eigenfunctions, we obtain the radial
equations for the spherically symmetric system, and analytically
solve the radial equations for the quantum Coulombic system.

This paper is organized as follows. Section 2 is devoted to the
generalization of the Dirac equation to $D+1$ space-time. In Sec.
3, the conserved angular momentum operators and their quantum
numbers are discussed. The eigenfunctions of the total angular
momentums are calculated for both odd $D$ and even $D$ cases in
terms of the method of group theory. The radial equations for the
system with a spherically symmetric potential are derived. In Sec.
4, the wave functions of bound states for a Coulombic system,
which are expressed by the confluent hypergeometric functions, are
presented. The energy levels and the corresponding fine structure
are also discussed. Some conclusions are given in Sec. 5.

\section{Dirac equation in $D+1$ dimensions}

The Dirac equation in $D+1$ dimensions can be expressed as \cite{bjo}
$$i \displaystyle \sum_{\mu=0}^{D}~\gamma^{\mu}\left(\partial_{\mu}
+ieA_{\mu}\right)\Psi({\bf x},t)=M\Psi({\bf x},t), \eqno (1) $$

\noindent

where $M$ is the mass of the particle, and (D+1) matrices
$\gamma_{\mu}$ satisfy the anti-commutative relations:
$$\gamma^{\mu}\gamma^{\nu}+\gamma^{\nu}\gamma^{\mu}=2\eta^{\mu \nu}{\bf 1},
\eqno (2) $$

\noindent

with the metric tensor $\eta^{\mu \nu}$ satisfying
$$\eta^{\mu \nu}=\eta_{\mu \nu}=\left\{\begin{array}{ll} \delta_{\mu \nu}
~~~~~~&{\rm when}~~\mu=0\\ -\delta_{\mu \nu} &{\rm when}~~\mu\neq
0.
\end{array} \right. \eqno (3) $$

\noindent

For simplicity, the natural units $\hbar=c=1$ are employed
throughout this paper. consider the special case where only the
zero component of $A_{\mu}$ is non-vanishing and spherically
symmetric:
$$\begin{array}{l}
eA_{0}=V(r),~~~~~~A_{a}=0,~~~~~~{\rm when}~~a\neq 0,  \end{array}
\eqno (4) $$

\noindent

The Hamiltonian $H({\bf x})$ of the system is expressed as
$$\begin{array}{l}
i\partial_{0}\Psi({\bf x},t)=H({\bf x})\Psi({\bf x},t),~~~~~~
H({\bf x})=\displaystyle \sum_{a=1}^{D}~\gamma^{0}\gamma^{a}p_{a}
+V(r)+\gamma^{0}M, \\
p_{a}=-i\partial_{a}=-i\displaystyle {\partial\over \partial
x^{a}}, ~~~~~~1\leq a \leq D. \end{array} \eqno (5) $$

The orbital angular momentum operators $L_{ab}$, the spinor
operators $S_{ab}$, and the total angular momentum operators
$J_{ab}$ are defined as follows
$$\begin{array}{ll}
L_{ab}=-L_{ba}=ix_{a}\partial_{b}-ix_{b}\partial_{a},~~~~~~
&S_{ab}=-S_{ba}=i\gamma_{a}\gamma_{b}/2,\\
J_{ab}=L_{ab}+S_{ab},~~~~~~&1\leq a <b \leq D\\
J^{2}=\displaystyle \sum_{a<b=2}^{D}~J_{ab}^{2},~~~~~~
L^{2}=\displaystyle \sum_{a<b=2}^{D}~L_{ab}^{2},~~~~
&S^{2}=\displaystyle \sum_{a<b=2}^{D}~S_{ab}^{2}. \end{array}
\eqno (6) $$

\noindent

The eigenvalue of $J^{2}$ ($L^{2}$ or $S^{2}$) is denoted by the
Casimir $C_{2}({\bf M})$, where ${\bf M}$ is the highest weight of
the representation to which the total (orbital or spinor) wave
function belongs. We will discuss the Casimir in the next section.
It is easy to show by the standard method \cite{bjo} that $J_{ab}$
and $\kappa$ are commutant with the Hamiltonian $H({\bf x})$,
$$\kappa=\gamma^{0}\left\{\displaystyle \sum_{a<b}~
i\gamma^{a}\gamma^{b}L_{ab}+(D-1)/2\right\}
=\gamma^{0}\left\{J^{2}-L^{2}-S^{2}+(D-1)/2\right\}. \eqno (7) $$

\section{The radial equations}

Since the potential $V(r)$ is spherically symmetric, the symmetry
group of the system is SO($D$) group. Erdelyi \cite {erd} and
Louck \cite{lou2,cha} introduced the hyperspherical coordinates in
the real $D$-dimensional space
$$\begin{array}{l}
x^{1}=r\cos \theta_{1} \sin \theta_{2} \ldots \sin \theta_{D-1},\\
x^{2}=r\sin \theta_{1} \sin \theta_{2} \ldots \sin \theta_{D-1}, \\
x^{b}=r\cos \theta_{b-1} \sin \theta_{k} \ldots \sin
\theta_{D-1},~~~~~~
3\leq b \leq D-1,\\
x^{D}=r\cos \theta_{D-1}\\
\displaystyle \sum_{a=1}^{D}~(x^{a})^{2}=r^{2}. \end{array} \eqno
(8) $$

\noindent

The unit vector along ${\bf x}$ is usually denoted by $\hat{\bf x}
={\bf x}/r$. The volume element of the configuration space is
$$\begin{array}{l}
\displaystyle \prod_{a=1}^{D} dx^{a}=r^{D-1}dr d\Omega,~~~~~~
d\Omega=\displaystyle
\prod_{a=1}^{D-1}\left(\sin~\theta_{a}\right)^{a-1}
d\theta_{a},\\
0\leq r \leq \infty,~~~~~~-\pi \leq \theta_{1} \leq \pi,~~~~~~
0\leq \theta_{b}\leq \pi,~~~~~~2\leq b \leq D-1. \end{array} \eqno
(9) $$

Now, let us sketch some necessary knowledge of the SO($D$) group.
From the representation theory of Lie groups \cite{fro,sal,ma2},
the Lie algebras of the SO(2$N$+1) group and the SO($2N$) group
are $B_{N}$ and $D_{N}$, respectively. Their Chevalley bases with
the subscript $\mu$, $1\leq \mu \leq N-1$, are same:

$$\begin{array}{l}
H_{\mu}(J)=J_{(2\mu-1)(2\mu)}-J_{(2\mu+1)(2\mu+2)},\\
E_{\mu}(J)=\left(J_{(2\mu)(2\mu+1)}-iJ_{(2\mu-1)(2\mu+1)}
-iJ_{(2\mu)(2\mu+2)}-J_{(2\mu-1)(2\mu+2)}\right)/2, \\
F_{\mu}(J)=\left(J_{(2\mu)(2\mu+1)}+iJ_{(2\mu-1)(2\mu+1)}
+iJ_{(2\mu)(2\mu+2)}-J_{(2\mu-1)(2\mu+2)}\right)/2. \end{array}
\eqno (10a) $$

\noindent

But, the bases with the subscript $N$ are different:
$$\begin{array}{l}
H_{N}(J)=2J_{(2N-1)(2N)},\\
E_{N}(J)=J_{(2N)(2N+1)}-iJ_{(2N-1)(2N+1)},\\
F_{N}(J)=J_{(2N)(2N+1)}+iJ_{(2N-1)(2N+1)},
\end{array} \eqno (10b) $$

\noindent

for SO($2N+1$), and
$$\begin{array}{l}
H_{N}(J)=J_{(2N-3)(2N-2)}+J_{(2N-1)(2N)}, \\
E_{N}(J)=\left(J_{(2N-2)(2N-1)}-iJ_{(2N-3)(2N-1)}
+iJ_{(2N-2)(2N)}+J_{(2N-3)(2N)}\right)/2, \\
F_{N}(J)=\left(J_{(2N-2)(2N-1)}+iJ_{(2N-3)(2N-1)}
-iJ_{(2N-2)(2N)}+J_{(2N-3)(2N)}\right)/2, \end{array} \eqno (10c)
$$

\noindent

for SO($2N$). The operator $J_{ab}$ can be replaced with $L_{ab}$
or $S_{ab}$ depending on the wave functions one is discussing.
$H_{\mu}(J)$ span the Cartan subalgebra, and their eigenvalues for
an eigenstate $|{\bf m}\rangle$ in a given irreducible
representation are the components of a weight vector ${\bf
m}=(m_{1},\ldots,m_{N})$:
$$H_{\mu}(J)|{\bf m}\rangle = m_{\mu}|{\bf m}\rangle,~~~~~~1\leq \mu \leq N.
\eqno (11) $$

\noindent

If the eigenstates $|{\bf m}\rangle$ for a given weight ${\bf m}$
are degenerate, this weight is called a multiple weight,
otherwise, a simple one. $E_{\mu}$ are called the raising
operators and $F_{\mu}$  the lowering ones. For an irreducible
representation, there is a  highest weight ${\bf M}$, which is a
simple weight and can be used to describe the irreducible
representation. Usually, the irreducible representation is also
called the highest weight representation and directly denoted by
${\bf M}$. The Casimir $C_{2}({\bf M})$ can be calculated by the
formula (e.g. see (1.131) in \cite{ma2})
$$C_{2}({\bf M})={\bf M}\cdot ({\bf M}+2{\bf \rho})
=\displaystyle \sum_{\mu, \nu=1}^{N}~M_{\mu}d_{\mu}(A^{-1})_{\mu
\nu} (M_{\nu}+2), \eqno (12) $$

\noindent

where ${\bf \rho}$ is the half sum of the positive roots in the
Lie algebra, $A^{-1}$ is the inverse of the Cartan matrix, and
$d_{\mu}$ are the half square lengths of the simple roots.

The orbital wave function in $D$-dimensional space is usually
expressed by the spherical harmonic $Y^{(l)}_{{\bf m}}(\hat{\bf
x})$ \cite{lou2,cha}, which belongs to the weight ${\bf m}$ of the
highest weight representation $(l)\equiv (l,0,\ldots,0)$. For the
highest weight state, ${\bf m}=(l)$, we have
$$\begin{array}{l}
Y^{(l)}_{(l)}(\hat{\bf x})=N_{D,l}r^{-l}(x^{1}+i x^{2})^{l},\\
N_{D,l}=\left\{\begin{array}{ll}2^{-N-l} \left\{\displaystyle
{(2l+2N-1)!\over \pi^{N} l!(l+N-1)!}\right\}^{1/2}
&{\rm when}~~D=2N+1\\
\left\{\displaystyle {(l+N-1)!\over 2 \pi^{N} l!}\right\}^{1/2}
&{\rm when}~~D=2N, \end{array}\right. \end{array} \eqno (13) $$

\noindent

where $N_{D,l}$ is the normalization factor. Its partners
$Y^{(l)}_{\bf m}(\hat{\bf x})$ can be calculated from
$Y^{(l)}_{(l)}(\hat{\bf x})$ by lowering operators $F_{\mu}(L)$.
The Casimir for the spherical harmonic $Y^{(l)}_{{\bf m}}(\hat{\bf
x})$ is calculated by Eq. (12):
$$L^{2}Y^{(l)}_{{\bf m}}(\hat{\bf x})=C_{2}[(l)]
Y^{(l)}_{{\bf m}}(\hat{\bf x}), ~~~~~~C_{2}[(l)]=l(l+D-2). \eqno
(14) $$

\noindent

The spinor wave functions as well as those for the total angular
momentum are different for $D=2N+1$ and $D=2N$, and will be
discussed separately in the following subsections.

\subsection{The SO($2N+1$) case}

For $D=2N+1$ we define
$$\gamma^{0}=\sigma_{3}\times {\bf 1},~~~~~~
\gamma^{a}=(i\sigma_{2})\times \beta_{a},~~~~~~1\leq a \leq 2N+1,
\eqno (15) $$

\noindent

where $\sigma_{a}$ is the Pauli matrix, ${\bf 1}$ denotes the
$2^{N}$-dimensional unit matrix, and $(2N+1)$ matrices $\beta_{a}$
satisfy the anticommutative relations
$$\beta_{a}\beta_{b}+\beta_{b}\beta_{a}=2\delta_{ab}{\bf 1},
~~~~~~a,~b=1,~2,\ldots,~(2N+1). \eqno (16) $$

\noindent

The dimension of $\beta_{a}$ matrices is $2^{N}$. Thus, the spinor
operator $S_{ab}$ becomes a block matrix
$$S_{ab}={\bf 1}\times \overline{S}_{ab},~~~~~~
\overline{S}_{ab}=-i\beta_{a}\beta_{b}/2. \eqno (17) $$

\noindent

The relation between $S_{ab}$ and $\overline{S}_{ab}$ is similar
to the relation between the spinor operators for the Dirac spinors
and for the Pauli spinors. The operator $\kappa$ becomes
$$\kappa=\sigma_{3}\times \overline{\kappa},~~~~~~
\overline{\kappa}=-i\displaystyle \sum_{a<b}
~\beta_{a}\beta_{b}L_{ab}+(D-1)/2. \eqno (18) $$

\noindent

The fundamental spinor $\chi({\bf m})$ belong to the fundamental
spinor representation $(s)\equiv (0,\ldots,0,1)$. From Eq. (12)
the Casimir for the representation $(s)$ is calculated to be
$C_{2}[(s)]=(2N^{2}+N)/4$.

The product of $Y^{(l)}_{{\bf m}}(\hat{\bf x})$ and $\chi({\bf
m'})$ belong to the direct product of two representation $(l)$ and
$(s)$, which is a reducible representation:
$$(l)\times (s)\simeq (l,0,\ldots,0,1)\oplus (l-1,0,\ldots,0,1).
\eqno (19) $$

\noindent

In other words, in order to construct a wave function belonging to
the representation $(j)\equiv (l,0,\ldots,0,1)$ there are two
different ways: the combination of $Y^{(l)}_{{\bf m}}(\hat{\bf
x})\chi({\bf m'})$ and that of $Y^{(l+1)}_{{\bf m}}(\hat{\bf
x})\chi({\bf m'})$. They have different eigenvalues of
$\overline{\kappa}$. Since the system is spherically symmetric, we
only need to calculate the highest weight state for the
representation $(j)$ in terms of the Clebsch-Gordan coefficients
$$\begin{array}{l}
\phi_{|K|,(j)}(\hat{\bf x}) =Y^{(l)}_{(l)}(\hat{\bf x})\chi[(s)]
=N_{D,l}r^{-l}(x^{1}+i x^{2})^{l}\chi[(s)], \\
|K|=C_{2}[(j)]-C_{2}[(l)]-C_{2}[(s)]+N =l+N. \end{array} \eqno
(20) $$

$$\begin{array}{l}
\phi_{-|K|,(j)}(\hat{\bf x}) =\displaystyle \sum_{\bf
m}~Y^{(l+1)}_{{\bf m}}(\hat{\bf x})
\chi[(j)-{\bf m}]\langle (l+1),{\bf m},(s),(j)-{\bf m}|(j),(j)\rangle\\
~~~=N_{D,l}r^{-l-1}(x^{1}+ix^{2})^{l}\left\{x^{2N+1}\chi[(s)]
+(x^{2N-1}+ix^{2N})\chi[(0,\ldots,0,1,\overline{1})]\right.\\
~~~~~~+(x^{2N-3}+ix^{2N-2})\chi[(0,\ldots,0,1,\overline{1},1)]+\ldots\\
~~~~~~\left.+(x^{3}+ix^{4})\chi[(1,\overline{1},0,\ldots,0,1)]
+(x^{1}+ix^{2})\chi[(\overline{1},0,\ldots,0,1)]\right\}, \\
-|K|=C_{2}[(j)]-C_{2}[(l+1)]-C_{2}[(s)]+N =-l-N. \end{array} \eqno
(21) $$

The wave function $\Psi_{K, (j)}({\bf x})$ of the total angular
momentum  belonging to the irreducible representation $(j)$ can be
expressed as
$$\begin{array}{l}
\Psi_{K, (j)}({\bf x},t)=r^{-N}e^{-iEt}\left(\begin{array}{c}
F(r)\phi_{K,(j)}(\hat{\bf x})\\
iG(r)\phi_{-K,(j)}(\hat{\bf x}) \end{array} \right), \\
H_{1}(J)\Psi_{K, (j)}({\bf x})=l \Psi_{K, (j)}({\bf x}),~~~~~~
H_{N}(J)\Psi_{K, (j)}({\bf x})=\Psi_{K, (j)}({\bf x}),\\
H_{\mu}(J)\Psi_{K, (j)}({\bf x})=0,~~~~~~2\leq \mu \leq N-1, \\
\kappa\Psi_{K, (j)}({\bf x})=K \Psi_{K, (j)}({\bf x}),~~~~~~ K=\pm
(l+N). \end{array} \eqno (22) $$

\noindent

Its partners can be calculated from it by the lowering operators
$F_{\mu}(J)$.

The radial equation will depend upon the explicit forms of
$\beta_{a}$ matrices. We express $\beta_{a}$ matrices by direct
products of $N$ Pauli matrices $\sigma_{a}$ \cite{geo}:
$$\begin{array}{l}
\beta_{2m-1}=\overbrace{{\bf 1}\times \ldots \times {\bf 1}}^{m-1}
\times \sigma_{1} \times \overbrace{\sigma_{3} \times \ldots
\times \sigma_{3}}^{N-m}, \\
\beta_{2m}=\overbrace{{\bf 1}\times \ldots \times {\bf 1}}^{m-1}
\times \sigma_{2} \times \overbrace{\sigma_{3} \times \ldots
\times \sigma_{3}}^{N-m}, \\
\beta_{2N+1}= \sigma_{3} \times \sigma_{3} \times \ldots \times
\sigma_{3}. \end{array} \eqno (23) $$

\noindent

In terms of the explicit forms of $\beta_{a}$, we obtain
$$\begin{array}{l}
\left(\vec{\beta}\cdot \hat{\bf x}\right)\phi_{K,(j)}(\hat{\bf x})
=r^{-1}\displaystyle \sum_{a=1}^{2N+1}~\beta_{a} x^{a}
~\phi_{K,(j)}(\hat{\bf x})=\phi_{-K,(j)}(\hat{\bf x}), \\
\left(\vec{\beta}\cdot \vec{\bf
p}\right)r^{-N}\phi_{K,(j)}(\hat{\bf x}) =\displaystyle
\sum_{a=1}^{2N+1}~\beta_{a} p_{a} ~r^{-N}\phi_{K,(j)}(\hat{\bf x})
=iKr^{-N-1}\phi_{-K,(j)}(\hat{\bf x}). \end{array} \eqno (24) $$

Substituting $\Psi_{K (j)}({\bf x})$ into the Dirac equation (5)
we obtain the radial equation
$$\begin{array}{l}
\displaystyle {dG(r) \over dr}+\displaystyle {K \over r}G(r)=
[E-V(r)-M] F(r), \\
-\displaystyle {dF(r) \over dr}+\displaystyle {K \over r}F(r)=
[E-V(r)+M] G(r). \end{array} \eqno (25) $$

\subsection{The SO($2N$) case}

As is well known, the spinor representation of SO($2N$) group is
reducible and can be reduced to two inequivalent fundamental
spinor representations $(+s)\equiv (0,0,\ldots,0,1)$ and
$(-s)\equiv (0,0,\ldots,0,1,0)$. From Eq. (12) the Casimir for
both spinor representations are calculated to be $C_{2}[(\pm
s)]=(2N^{2}-N)/4$. In terms of $\beta_{a}$ matrices given in Eq.
(23), we define $\gamma^{\mu}$ matrices for $D=2N$:
$$\gamma^{0}=\beta_{2N+1},~~~~~~
\gamma^{a}=\beta_{2N+1} \beta_{a},~~~~~~1\leq a \leq 2N.  \eqno
(26) $$

\noindent $\gamma^{0}$ is a diagonal matrix where half of the
diagonal elements are equal to $+1$ and the remaining to $-1$.
Because the spinor operator $S_{ab}$ and the operator $\kappa$ are
commutant with $\gamma^{0}$, each of them becomes a direct sum of
two matrices, referring to the rows with the eigenvalues $+1$ and
$-1$ of $\gamma^{0}$, respectively. The fundamental spinors
$\chi_{\pm}({\bf m})$ belong to the fundamental spinor
representations $(+s)$ and $(-s)$, respectively, and satisfy
$$\gamma^{0}\chi_{\pm}({\bf m})=\pm \chi_{\pm}({\bf m}). \eqno (27) $$

The product of $Y^{(l)}_{{\bf m}}(\hat{\bf x})$ and
$\chi_{\pm}({\bf m'})$ belong to the direct product of two
representation $(l)$ and $(\pm s)$, which is a reducible
representation:

$$\begin{array}{l}
(l)\times (+s)\simeq (l,0,\ldots,0,1)\oplus
(l-1,0,\ldots,0,1,0), \\
(l)\times (-s)\simeq (l,0,\ldots,0,1,0)\oplus (l-1,0,\ldots,0,1).
\end{array} \eqno (28) $$

\noindent

There are two kinds of representations for the total angular
momentum: the representation $(j_{1})\equiv (l,0,\ldots,0,1)$ and
the representation $(j_{2})\equiv (l,0,\ldots,0,1,0)$. Their
Casimirs are the same:
$$C_{2}[(j_{1})]=C_{2}[(j_{2})]
=l(l+2N-1)+(2N^{2}-N)/4. \eqno (29) $$

There are two different ways to construct a wave function
belonging to the representation $(j_{1})$: the combination of
$Y^{(l)}_{{\bf m}}(\hat{\bf x})\chi_{+}({\bf m'})$ and that of
$Y^{(l+1)}_{{\bf m}}(\hat{\bf x})\chi_{-}({\bf m'})$. Due to the
spherical symmetry, we only calculate the highest weight state for
the representation $(j_{1})$ by the Clebsch-Gordan coefficients:
$$\begin{array}{l}
\phi_{K,(j_{1})}(\hat{\bf x}) =Y^{(l)}_{(l)}(\hat{\bf
x})\chi_{+}[(+s)]
=N_{D,l}r^{-l}(x^{1}+i x^{2})^{l}\chi_{+}[(+s)], \\
\phi_{-K,(j_{1})}(\hat{\bf x}) =\displaystyle \sum_{\bf
m}~Y^{(l+1)}_{{\bf m}}(\hat{\bf x}) \chi_{-}[(j_{1})-{\bf
m}]\langle (l+1),{\bf m},(-s),(j_{1})-{\bf m}|
(j_{1}),(j_{1})\rangle\\
~~~=N_{D,l}r^{-l-1}(x^{1}+ix^{2})^{l}\left\{(x^{2N-1}+ix^{2N})\chi_{-}[(-s)]
\right.\\
~~~~~~+(x^{2N-3}+ix^{2N-2})\chi_{-}[(0,\ldots,0,1,\overline{1},0)]\\
~~~~~~+(x^{2N-5}+ix^{2N-4})\chi_{-}[(0,\ldots,0,1,\overline{1},0,1)]+\ldots\\
~~~~~~\left.+(x^{3}+ix^{4})\chi_{-}[(1,\overline{1},0,\ldots,0,1)]
+(x^{1}+ix^{2})\chi_{-}[(\overline{1},0,\ldots,0,1)]\right\}, \\
K=C_{2}[(j_{1})]-C_{2}[(l)]-C_{2}[(+ s)]+N-1/2 =l+N-1/2.
\end{array} \eqno (30) $$

\noindent

For the representation $(j_{2})\equiv (l,0,\ldots,0,1,0)$ we have
$$\begin{array}{l}
\phi_{K,(j_{2})}(\hat{\bf x}) =\displaystyle \sum_{\bf
m}~Y^{(l+1)}_{{\bf m}}(\hat{\bf x}) \chi_{+}[(j_{2})-{\bf
m}]\langle (l+1),{\bf m},(+s),(j_{2})-{\bf m}|
(j_{2}),(j_{2})\rangle\\
~~~=N_{D,l}r^{-l-1}(x^{1}+ix^{2})^{l}\left\{(x^{2N-1}-ix^{2N})\chi_{+}[(+s)]
\right.\\
~~~~~~+(x^{2N-3}+ix^{2N-2})\chi_{+}[(0,\ldots,0,1,0,\overline{1})]\\
~~~~~~+(x^{2N-5}+ix^{2N-4})\chi_{+}[(0,\ldots,0,1,\overline{1},1,0)]+\ldots\\
~~~~~~\left.+(x^{3}+ix^{4})\chi_{+}[(1,\overline{1},0,\ldots,0,1,0)]
+(x^{1}+ix^{2})\chi_{+}[(\overline{1},0,\ldots,0,1,0)]\right\}, \\
\phi_{-K,(j_{2})}(\hat{\bf x}) =Y^{(l)}_{(l)}(\hat{\bf
x})\chi_{-}[(-s)]
=N_{D,l}r^{-l}(x^{1}+i x^{2})^{l}\chi_{-}[(-s)], \\
K=C_{2}[(j_{2})]-C_{2}[(l+1)]-C_{2}[(+s)]+N-1/2 =-l-N+1/2.
\end{array} \eqno (31) $$

\noindent
In terms of the explicit forms of $\beta_{a}$ we obtain
$$\begin{array}{l}
\left(\vec{\beta}\cdot \hat{\bf
x}\right)\phi_{K,(j_{\omega})}(\hat{\bf x}) =r^{-1}\displaystyle
\sum_{a=1}^{2N}~\beta_{a} x^{a} \phi_{K,(j_{\omega})}(\hat{\bf x})
=\phi_{-K,(j_{\omega})}(\hat{\bf x}), \\
\left(\vec{\beta}\cdot \vec{\bf p}\right)
r^{-N+1/2}\phi_{K,(j_{\omega})}(\hat{\bf x}) =\displaystyle
\sum_{a=1}^{2N}~\beta_{a} p_{a}
~r^{-N+1/2}\phi_{K,(j_{\omega})}(\hat{\bf x})
=iKr^{-N-1/2}\phi_{-K,(j_{\omega})}(\hat{\bf x})\\
\omega=1~~{\rm or}~~2.
\end{array} \eqno (32) $$

The wave function $\Psi_{K,(j_{\omega})}({\bf x})$ of the total
angular momentum belonging to the irreducible representation
$(j_{\omega})$ can be expressed as
$$\begin{array}{l}
\Psi_{|K|, (j_{1})}({\bf x},t)=r^{-N+1/2}e^{-iEt}\left\{
F(r)\phi_{|K|,(j_{1})}(\hat{\bf x})+i
G(r)\phi_{-|K|, (j_{1})}(\hat{\bf x}) \right\}, \\
\Psi_{-|K|, (j_{2})}({\bf x},t)=r^{-N+1/2}e^{-iEt}\left\{
F(r)\phi_{-|K|,(j_{2})}(\hat{\bf x})+i
G(r)\phi_{|K|, (j_{2})}(\hat{\bf x}) \right\}, \\
\kappa\Psi_{K, (j_{\omega})}({\bf x})=K \Psi_{K,
(j_{\omega})}({\bf x}), ~~~~~~K=\left\{\begin{array}{ll}
 l+N-1/2,~~~~~~&{\rm when}~~\omega=1\\
-l-N+1/2,~~~~~~&{\rm when}~~\omega=2, \end{array} \right. \\
H_{1}(J)\Psi_{K, (j_{\omega})}({\bf x})=l \Psi_{K, (j_{\omega})}({\bf x}),\\
H_{N-1}(J)\Psi_{K, (j_{1})}({\bf x})=0,~~~~~~
H_{N}(J)\Psi_{K, (j_{1})}({\bf x})=\Psi_{K, (j_{1})}({\bf x}),\\
H_{N-1}(J)\Psi_{K, (j_{2})}({\bf x})=\Psi_{K, (j_{2})}({\bf
x}),~~~~~~
H_{N}(J)\Psi_{K, (j_{2})}({\bf x})=0,\\
H_{\mu}(J)\Psi_{K, (j_{\omega})}({\bf x})=0,~~~~~~2\leq \mu \leq
N-2.
\end{array} \eqno (33) $$

\noindent

Its partners can be calculated from it by the lowering operators
$F_{\mu}(J)$. Substituting $\Psi_{K (j_{\omega})}({\bf x})$ into
the Dirac equation (5) we obtain the radial equations, which are
in the same forms as those in $D=2N+1$ case:
$$\begin{array}{l}
\displaystyle {dG(r) \over dr}+\displaystyle {K \over r}G(r)=
[E-V(r)-M] F(r), \\
-\displaystyle {dF(r) \over dr}+\displaystyle {K \over r}F(r)=
[E-V(r)+M] G(r). \end{array} \eqno (34) $$

\section {Solutions to the radial equation in D+1 dimensions}

Although the wave functions and the eigenvalues $K$ are different
for the $D=2N+1$ case and the $D=2N$ case, the forms of the radial
equations are unified
$$\begin{array}{l}
\displaystyle {dG_{KE}(r) \over dr}+\displaystyle {K\over
r}G_{KE}(r)=
[E-V(r)-M] F_{KE}(r), \\
-\displaystyle {dF_{KE}(r) \over dr}+\displaystyle {K\over
r}F_{KE}(r)=
[E-V(r)+M] G_{KE}(r), \\
K=\pm (2l+D-1)/2. \end{array} \eqno (35) $$

\noindent

For definiteness we discuss the attractive Coulomb potential
$$V(r)=-\displaystyle {\xi \over r},~~~~~~\xi=Z \alpha>0, \eqno (36) $$

\noindent

where $\alpha=1/137$ is the fine structure constant. It is easy to
see that the solution for the repulsive potential can be obtained
from that for the attractive potential by interchanging
$$F_{KE}\longleftrightarrow G_{-K-E},~~~~~~V(r)\longleftrightarrow -V(r).
\eqno (37) $$

\noindent

From the Sturm-Liouville theorem \cite{dong2}, there are bound
states with the energy less than and near $M$ for the attractive
Coulomb potential and with the energy larger than and near $-M$
for the repulsive potential, if the interaction is not too strong.
It is convenient to introduce a dimensionless variable $\rho$ in
Eq. (35) for bound states:
$$\rho=2r\sqrt{M^{2}-E^{2}},~~~~~~0<E< M. \eqno (38) $$

\noindent

Solving $F(\rho)$ from Eq. (35),
$$F_{KE}(\rho)=\left(-\displaystyle {1\over 2}
\sqrt{\displaystyle {M-E \over M+E}} +\displaystyle {\xi \over
\rho}\right)^{-1} \left[\displaystyle {d G_{KE}(\rho) \over
d\rho}+ \displaystyle {K\over \rho}G_{KE}(\rho) \right], \eqno
(39) $$

\noindent

we obtain a second-order differential equation of $G_{KE}(\rho)$:
$$\begin{array}{l}
\displaystyle {d^{2} G_{KE}(\rho) \over d\rho^{2}}
+\left[-\displaystyle {1\over 4} -\displaystyle
{K^{2}-\xi^{2}+K\over \rho^{2}}
+\displaystyle {E \xi \over \rho \sqrt{M^{2}-E^{2}}}\right]G_{KE}(\rho)\\
~~~~~~+\left[\rho-\displaystyle {\rho^{2}\over 2\xi}
\sqrt{\displaystyle {M-E \over M+E}}\right]^{-1}
\left[\displaystyle {d G_{KE}(\rho) \over d\rho}+ \displaystyle
{K\over \rho}G_{KE}(\rho) \right]=0. \end{array} $$

\noindent

From the behavior of $G_{KE}(\rho)$ at the origin and at the
infinity, we define
$$\begin{array}{l}
G_{KE}(\rho)=\rho^{\lambda}e^{-\rho/2}R(\rho),~~~~~~
\lambda=\sqrt{K^{2}-\xi^{2}}>0,\\
\omega=\displaystyle {1\over 2\xi} \sqrt{\displaystyle {M-E \over
M+E}},~~~~~~ \tau=\displaystyle {E \xi \over \sqrt{M^{2}-E^{2}}},
\end{array} \eqno (40) $$

\noindent

and obtain
$$\begin{array}{l}
(\rho-\omega \rho^{2})\displaystyle {d^{2} R(\rho) \over
d\rho^{2}}+\left[\omega\rho^{2}-\left(2\lambda\omega+1\right)
\rho+2\lambda+1\right] \displaystyle {d R(\rho) \over d\rho }\\
~~~~~~+\left[\omega(\lambda-\tau)\rho+\omega(K+\lambda)
+\tau-\lambda-1/2\right]R(\rho)=0.
\end{array} \eqno (41) $$

Eq. (41) can be solved by the power series expansion method for
for (3+1)-dimensions \cite{wal} and (2+1)-dimensions \cite{don}.
The results in $D+1$ dimensions are as follows
$$\begin{array}{rl}
\left. \begin{array}{l} F_{KE}(\rho)\\
G_{KE}(\rho)\end{array}\right\} &=~\displaystyle
{(M^{2}-E^{2})^{1/4} \over \Gamma(2\lambda+1)} \sqrt{\displaystyle
{(M\pm E) E \Gamma(n'+2\lambda+1) \over
2M^{2}\tau(K+\tau M/E)n'!}}~\rho^{\lambda} e^{-\rho/2}\\
&~~~\times \left[(K+\tau M/E)~_{1}F_{1}(-n',2\lambda+1,\rho) \mp
n'~_{1}F_{1}(1-n',2\lambda+1,\rho)\right], \end{array} \eqno (42)
$$

$$\displaystyle \int_{0}^{\infty}\left(|F_{KE}(\rho)|^{2}
+|G_{KE}(\rho)|^{2}\right)dr=1,  \eqno (43) $$
$$n^{\prime}=\tau-\lambda=0, 1, 2, \ldots .\eqno (44) $$

\noindent

where $_{1}F_{1}(\alpha,\beta,\rho)$ is the confluent
hypergeometric function. When $n^{\prime}=0$, $K$ has to be
positive. Introduce the principal quantum number
$$n=|K|-(D-3)/2+n^{\prime}=|K|-(D-3)/2+\tau-\lambda
=1,~2,~\ldots . \eqno (45) $$

\noindent

The principal quantum number $n$ can be equal to 1 only for
$K=(D-1)/2$ and equal to other positive integers for both signs of
$K$. The energy $E$ can be calculated from Eqs. (40), (44) and
(45)
$$E=M\left[1+\frac{\xi^2}{(\sqrt{K^2-\xi^2}+n-|K|+(D-3)/2)^2}
\right]^{-1/2}. \eqno (46) $$

\noindent

Expanding Eq. (46) in powers of $\xi^{2}$, we have
$$E\simeq M\left\{1-\displaystyle {\xi^2\over 2[n+(D-3)/2]^{2}}
-\displaystyle {\xi^4 \over 2[n+(D-3)/2]^{4}}\left( \displaystyle
{n+(D-3)/2 \over |K|}-\displaystyle {3\over 4}\right)\right\},
\eqno (47) $$

\noindent

where the first term on the right hand side is the rest energy $M$
($c^2=1$ in our conventions), the second one coincides with the
energy from the solutions to the Schr\"{o}dinger equation, and the
third one is the fine structure energy, which removes the
degeneracy between the states of the same $n$.

\section{Conclusions}

In this paper we generalized the Dirac equation to
(D+1)-dimensional space-time. The conserved angular momentum
operators and their quantum numbers are discussed. The
eigenfunctions of the total angular momentums are calculated for
both odd $D$ and even $D$ cases, respectively. The unified radial
equations for a spherically symmetric system are obtained. The
radial equations with a Coulomb potential are solved by the power
series expansion approach. The exact solutions are expressed by
the confluent hypergeometric functions. The eigenvalues as well as
their fine structure energy are also studied. Our solutions
coincide with those in 3+1 dimensional \cite{wal} and 2+1
dimensional \cite{don} space-time.

\vspace{5mm}

\noindent

{\bf ACKNOWLEDGMENTS} S.-H. Dong would like to thank Professor A.
Frank for the hospitality in UNAM. This work is supported by the
National Natural Science Foundation of China and CONACyT, Mexico,
under project 32397-E.

\end{document}